\documentclass[aip,jcp,reprint,groupedaddress,floatfix]{revtex4-1}

\usepackage{amsmath}
\usepackage{amssymb}
\usepackage{graphicx}
\usepackage{mathtools}
\usepackage[version=4]{mhchem}
\usepackage{microtype}
\usepackage{physics}
\usepackage{siunitx}

\usepackage[hidelinks]{hyperref}

\DeclareMathOperator{\sinc}{sinc}

\renewcommand{\vec}[1]{\boldsymbol{\mathbf{#1}}}
\newcommand{\tilvec}[1]{\tilde{\mathbf{#1}}}

\renewcommand{\phi}{\varphi}

\newcommand{\kB}{k_\mathrm{B}}

\newcommand{\trans}{^{\mathrm{T}}}
\newcommand{\inv}{^{-1}}
\newcommand{\st}{^\star}
\newcommand{\cl}{_\mathrm{cl}}
\newcommand{\constr}{^\mathrm{c}}
\newcommand{\ddf}[1]{\delta\qty(#1)}
\newcommand{\dlogJ}[1]{\pdv{\xi} \log{\abs*{J(#1)}}}

\newcommand{\estim}{\mathcal{E}}
\newcommand{\dt}{\Delta t}

\newcommand{\sep}{\,;\,\,}

\begin{document}

\title{
	On the quantum mechanical potential of mean force. II. Constrained path integral molecular dynamics integrators
}

\author{Dmitri Iouchtchenko}
\author{Kevin P. Bishop}
\author{Pierre-Nicholas Roy}
\email{pnroy@uwaterloo.ca}
\affiliation{Department of Chemistry, University of Waterloo, Waterloo, Ontario, N2L 3G1, Canada}

\begin{abstract}
Building on Paper I of this series, which introduced path integral Monte Carlo (PIMC) estimators for the derivative of the potential of mean force (PMF), we propose two path integral molecular dynamics (PIMD) integrators that can make use of these estimators.
These integrators, c-OBABO and c-BAOAB, are based on the path integral Langevin equation (PILE) integrator, which has seen widespread success in PIMD applications, but they include support for holonomic constraints.
When the reaction coordinate is the distance between two centers of mass, we find that several exact expressions are accessible: the Fixman correction, the position constraint Lagrange multiplier, and various derivatives with respect to the reaction coordinate.
It is observed that c-BAOAB tends to have a smaller time step error than c-OBABO, which is consistent with previous studies on integrator step ordering in molecular dynamics with holonomic constraints and in PIMD.
Further, we show that both the PMF of a water dimer and its derivative may be obtained from PIMD simulations using c-BAOAB, yielding results in agreement with the path integral umbrella sampling method previously used for this system.
\end{abstract}

\maketitle

\section{Introduction}
\label{sec:introduction}

In Paper I of this series, we demonstrated that the quantum potential of mean force (PMF) may be obtained from constrained path integral Monte Carlo (PIMC) simulations using novel estimators for the derivative of the PMF.
However, devising efficient configuration update schemes for Monte Carlo methods is a nontrivial endeavor, as the volume of configuration space grows exponentially both with the number of particles and the number of beads per particle.
For most systems, it is not feasible to simply perturb the position of each bead by a random amount, since the magnitude of the perturbations must be decreased substantially when increasing the number of beads, in order to avoid rejecting all proposed configurations.
Thus, it is often beneficial to turn to path integral molecular dynamics (PIMD), in which configurations are updated based on the force arising from an effective Hamiltonian.
In the unconstrained case, the path integral Langevin equation (PILE) integrator provides a straightforward implementation of thermostatted ring polymer time evolution for PIMD.\cite{ceriotti2010efficient}

In the present work, we incorporate bead-local holonomic constraints into the PILE in order to calculate the quantum PMF of molecular systems.
We also show that when the reaction coordinate for the PMF is the radial separation between two centers of mass, the additional computational effort necessary to enforce the constraint is negligible, as the Lagrange multiplier may be computed directly without resorting to an iterative scheme.

A recent study has found that an approach which combines PIMD with umbrella sampling and histogram unbiasing is sufficient to calculate the quantum PMF of a water dimer.\cite{bishop2018quantum}
While it may be advantageous to stitch together multiple histograms, as each simulation will contribute data over a range of the reaction coordinate, it can be quite challenging to obtain and converge a collection of smooth histograms.
On the other hand, independent values obtained from an estimator at single points are simple to examine and improve in a systematic fashion.
Indeed, we observe that when the PMF has a very rapidly changing slope, as in the water dimer at low temperature, it can be favorable to compute its derivative using constrained PIMD and our estimators rather than by numerical differentiation of an unbiased PMF.

The remainder of this article is organized as follows: in Sec.~\ref{sec:background}, we describe our notation; in Sec.~\ref{sec:integrators}, we develop two integrators for constrained path integral Langevin dynamics; in Sec.~\ref{sec:exact-constraint}, we derive exact expressions for a special case of the reaction coordinate; in Sec.~\ref{sec:results}, we apply the integrators to a water dimer; and in Sec.~\ref{sec:conclusions}, we summarize our findings.

\section{Background}
\label{sec:background}

As in Paper I, we consider Hamiltonians of the form
\begin{align}
	\hat{H}
	&= \frac{1}{2} \hat{\tilvec{p}} \cdot \tilvec{M}\inv \cdot \hat{\tilvec{p}} + V(\hat{\tilvec{q}})
\end{align}
(note the addition of tildes to distinguish these quantities from the fictitious ones in the molecular dynamics simulations), and work with the classical potential
\begin{align}
	V\cl(\vec{q})
	&= \sum_{i=1}^f \frac{\tilde{m}_i P}{2 \hbar^2 \tilde{\beta}^2} \sum_{j=1}^P \left( q^{(j)}_i - q^{(j+1)}_i \right)^2
		+ \frac{1}{P} \sum_{j=1}^P V(\vec{q}^{(j)}),
\end{align}
which is extracted from the discrete imaginary time path integral of the quantum partition function $Z = \Tr e^{-\tilde{\beta} \hat{H}}$ at reciprocal temperature $\tilde{\beta}$ with $P$ beads.

To sample from the unconstrained path density $e^{-\tilde{\beta} V\cl(\vec{q})}$ using molecular dynamics, one may associate a fictitious mass $m^{(j)}_i$ and momentum $p^{(j)}_i$ with each Cartesian bead coordinate $q^{(j)}_i$.\cite{parrinello1984study}
The masses form the $P f \times P f$ diagonal mass matrix $\vec{M}$, while the momenta are collected into the vector $\vec{p}$ of length $P f$.
We require that all the fictitious masses for a single degree of freedom be equal ($m^{(j)}_i = m_i$), so that each corresponding block of $\vec{M}$ is guaranteed to be invariant under every similarity transformation.
We consider only the typical case when the ratio $\tilde{m}_i / m_i$ is the same for all degrees of freedom; we refer to said ratio as $\tilde{m} / m$ and define the single-bead fictitious mass matrix as
\begin{align}
	\vec{M}_1
	&= \frac{m}{\tilde{m}} \tilde{\vec{M}}.
\end{align}

For a simulation at a fictitious reciprocal temperature $\beta = 1 / \kB T$, the momentum partition function
\begin{align}
	\label{eq:Z-p}
	\int\! \dd{\vec{p}} \, e^{-\frac{\beta}{2} \vec{p} \cdot \vec{M}\inv \cdot \vec{p}}
	&= \left( \frac{2 \pi}{\beta} \right)^{\frac{P f}{2}} \abs{\vec{M}}^\frac{1}{2}
\end{align}
may be inserted into configurational averages to express them as phase space averages:
\begin{align}
	\frac{
			\int\! \dd{\vec{q}} \, e^{-\tilde{\beta} V\cl(\vec{q})} f(\vec{q})
		}{
			\int\! \dd{\vec{q}} \, e^{-\tilde{\beta} V\cl(\vec{q})}
		}
	&= \frac{
			\int\! \dd{\vec{p}} \int\! \dd{\vec{q}} \, e^{-\beta H\cl(\vec{p}, \vec{q})} f(\vec{q})
		}{
			\int\! \dd{\vec{p}} \int\! \dd{\vec{q}} \, e^{-\beta H\cl(\vec{p}, \vec{q})}
		},
\end{align}
with the classical Hamiltonian given by
\begin{align}
	\label{eq:H-cl}
	H\cl(\vec{p}, \vec{q})
	&= \frac{1}{2} \vec{p} \cdot \vec{M}\inv \cdot \vec{p} + \frac{\tilde{\beta}}{\beta} V\cl(\vec{q}).
\end{align}
Although it appears that we've increased the complexity of the integrals, the phase space formulation of the expectation value is readily evaluated using molecular dynamics techniques.

\subsection{Path integral Langevin equation (PILE) integrator}

The PILE integrator is a combination of a white noise Langevin thermostat with a generalized velocity Verlet scheme that operates on path normal modes.\cite{ceriotti2010efficient}
The normal mode transformation is done via an orthogonal matrix $C$ with elements
\begin{align}
	C_{j k}
	&= \begin{dcases*}
			\sqrt{\frac{1}{P}} & if $k = 0$ \\
			\sqrt{\frac{2}{P}} \cos{\frac{2\pi j k}{P}} & if $1 \le k < \frac{P}{2}$ \\
			\sqrt{\frac{1}{P}} (-1)^j & if $k = \frac{P}{2}$ \\
			\sqrt{\frac{2}{P}} \sin{\frac{2\pi j k}{P}} & if $\frac{P}{2} < k \le P - 1$,
		\end{dcases*}
\end{align}
where $j$ labels the beads ($1$ to $P$) and $k$ labels the modes ($0$ to $P-1$).
We use the vectors $\vec{P}$ and $\vec{Q}$ for the transformed coordinates.

This integrator is composed of three exact sub-integrators.
The propagation of harmonic oscillators in normal mode coordinates is
{\small
\begin{align}
	\label{eq:subint-ho}
	\rotatebox[origin=c]{90}{A} &\begin{dcases}
		\vec{P}^{(k)}
		\leftarrow \textstyle\sum_{j=1}^P \vec{p}^{(j)} C_{j k} \sep
		\vec{Q}^{(k)}
		\leftarrow \textstyle\sum_{j=1}^P \vec{q}^{(j)} C_{j k} \\[1mm]
		\bar{\vec{P}}^{(k)}
		\leftarrow S^{\vec{P} \vec{P}}_k(\dt)               \vec{P}^{(k)} + S^{\vec{P} \vec{Q}}_k(\dt) \vec{M}_1 \vec{Q}^{(k)} \\
		\bar{\vec{Q}}^{(k)}
		\leftarrow S^{\vec{Q} \vec{P}}_k(\dt) \vec{M}_1\inv \vec{P}^{(k)} + S^{\vec{Q} \vec{Q}}_k(\dt)           \vec{Q}^{(k)} \\[1mm]
		\bar{\vec{p}}^{(j)}
		\leftarrow \textstyle\sum_{k=0}^{P-1} C_{j k} \bar{\vec{P}}^{(k)} \sep
		\bar{\vec{q}}^{(j)}
		\leftarrow \textstyle\sum_{k=0}^{P-1} C_{j k} \bar{\vec{Q}}^{(k)},
	\end{dcases}
\end{align}
}%
where we have made explicit the transformations to and from normal modes.
The propagation coefficients
\begin{subequations}
\begin{align}
	S^{\vec{P} \vec{P}}_k(\dt)
	= S^{\vec{Q} \vec{Q}}_k(\dt)
	&= \cos(\omega_k \dt), \\
	S^{\vec{P} \vec{Q}}_k(\dt)
	&= -\omega_k \sin(\omega_k \dt), \\
\intertext{and}
	S^{\vec{Q} \vec{P}}_k(\dt)
	&= \dt \sinc\qty(\omega_k \dt)
\end{align}
\end{subequations}
arise from the exact solution of Hamilton's equations of motion for a harmonic oscillator with angular frequency
\begin{align}
	\omega_k
	&= 2 \sqrt{\frac{\tilde{m} P}{\hbar^2 m \tilde{\beta} \beta}} \sin{\frac{\pi k}{P}}.
\end{align}

Application of the remaining force may be done in Cartesian coordinates:
{\small
\begin{align}
	\rotatebox[origin=c]{90}{B} &\begin{dcases}
		\bar{\vec{p}}^{(j)}
		\leftarrow \vec{p}^{(j)} + \frac{\dt \tilde{\beta}}{P \beta} \vec{F}(\vec{q}^{(j)}),
	\end{dcases}
\end{align}
}%
where
\begin{align}
	\vec{F}(\vec{q}^{(j)})
	&= -\grad V(\vec{q}^{(j)}).
\end{align}

The normal mode degrees of freedom are independently thermostatted using a Langevin thermostat:
{\small
\begin{align}
	\rotatebox[origin=c]{90}{O} &\begin{dcases}
		\vec{P}^{(k)}
		\leftarrow \textstyle\sum_{j=1}^P \vec{p}^{(j)} C_{j k} \\[1mm]
		\bar{\vec{P}}^{(k)}
		\leftarrow T_k(\dt) \vec{P}^{(k)} + U_k(\dt) \vec{M}_1^{\frac{1}{2}} \vec{\eta} \\[1 mm]
		\bar{\vec{p}}^{(j)}
		\leftarrow \textstyle\sum_{k=0}^{P-1} C_{j k} \bar{\vec{P}}^{(k)},
	\end{dcases}
\end{align}
}%
where the coefficients are
\begin{subequations}
\begin{align}
	T_k(\dt)
	&= e^{-\dt \gamma_k}
\intertext{and}
	U_k(\dt)
	&= \sqrt{\frac{1}{\beta} (1 - e^{-2 \dt \gamma_k})},
\end{align}
\end{subequations}
$\gamma_k$ is a friction coefficient (the same for all degrees of freedom), and $\vec{\eta}$ is a vector of $f$ pseudorandom numbers sampled from a standard normal distribution.
When $k \ge 1$, the value of $\gamma_k$ that minimizes the energy autocorrelation time in the free ($V(\tilvec{q}) = 0$) case is known analytically to be $2 \omega_k$, and this value is typically used even when interactions are present; the centroid friction $\gamma_0$ is a tunable simulation parameter.

If these sub-integrators are referred to as A, B, and O, respectively, then they may be combined in the order OBABO to form the PILE integrator.
It is implied by this notation that all the steps other than the central (that is, both repetitions of B and O) have a halved duration of $\dt / 2$, as required by the symmetric splitting of the Fokker--Planck operator.

\subsection{Constrained Hamiltonian integrators}
\label{sec:constrained-hamiltonian}

The addition of holonomic constraints to a symplectic integrator for Hamilton's equations of motion may be accomplished by a straightforward scheme, in which non-dynamical momentum perturbations ensure that both the position and velocity constraints are satisfied at the end of each step.\cite{reich1996symplectic}
For example, this may be used to obtain the well-known RATTLE algorithm\cite{andersen1983rattle} from velocity Verlet.

In this scheme, a generic integration step of the form
{\small
\begin{align}
	&\begin{dcases}
		\bar{\vec{p}}
		\leftarrow f_{\vec{p}}(\vec{p}, \vec{q}) \\
		\bar{\vec{q}}
		\leftarrow f_{\vec{q}}(\vec{p}, \vec{q})
	\end{dcases}
\end{align}
}%
becomes the two-step sequence
\begin{subequations}\small
\begin{align}
	&\begin{dcases}
		\vec{p}'
		\leftarrow \vec{p} + \grad \vec{\xi}(\vec{q}) \cdot \vec{\Lambda} \\
		\bar{\vec{p}}
		\leftarrow f_{\vec{p}}(\vec{p}', \vec{q}) \\
		\bar{\vec{q}}
		\leftarrow f_{\vec{q}}(\vec{p}', \vec{q}) \\
		\vec{\xi}(\bar{\vec{q}}) = \vec{z}
	\end{dcases} \\
	&\begin{dcases}
		\bar{\vec{p}}
		\leftarrow \vec{p} + \grad \vec{\xi}(\vec{q}) \cdot \vec{\Lambda} \\
		\dot{\vec{\xi}}(\vec{q}) = 0,
	\end{dcases}
\end{align}
\end{subequations}
where $\vec{\xi}(\vec{q}) = \vec{z}$ is the holonomic constraint to be maintained, $\dot{\vec{\xi}}(\vec{q}) = 0$ is the implicit velocity constraint, and each $\vec{\Lambda}$ is a vector of Lagrange multipliers that results in the final line of the corresponding step being valid.
Note how the first step begins by perturbing the momentum away from the constraint manifold in order to ensure that the position constraint is satisfied, while the second step projects the momentum back onto the constraint manifold.

In principle, the Lagrange multipliers may be found at each step of the simulation by integrating their equations of motion explicitly.
However, this will result in a growing discrepancy between the calculated values and the values that are needed to correctly enforce the constraints.\cite{ryckaert1977numerical}
Instead, one should solve a system of equations for the Lagrange multipliers at each step; since these equations are generally nonlinear, they are most often solved by iteration.

\section{Integrators}
\label{sec:integrators}

Our aim is to compute the derivative of the PMF,
\begin{align}
	-\tilde{\beta} A'(\xi\st)
	= \frac{
			\int\! \dd{\vec{q}} \, \ddf{\xi(\vec{q}^{(1)}) - \xi\st} e^{-\tilde{\beta} V\cl(\vec{q})} \estim_i(\vec{q})
		}{
			\int\! \dd{\vec{q}} \, \ddf{\xi(\vec{q}^{(1)}) - \xi\st} e^{-\tilde{\beta} V\cl(\vec{q})}
		},
\end{align}
in a PIMD setting using the two path integral estimators
\begin{subequations}
\begin{align}
	\estim_1(\vec{q})
	&= \dlogJ{\vec{q}^{(1)}}
		+ \frac{\tilde{\beta}}{P} \sum_{j=1}^P \vec{F}(\vec{q}^{(j)}) \cdot \pdv{\vec{q}^{(1)}}{\xi}
\intertext{and}
	\estim_2(\vec{q})
	&= \dlogJ{\vec{q}^{(1)}}
		+ \tilde{\beta} \vec{F}\cl^{(1)}(\vec{q}) \cdot \pdv{\vec{q}^{(1)}}{\xi}
\end{align}
\end{subequations}
from Paper I.
Although the molecular dynamics simulations will take place in Cartesian coordinates, it is substantially more convenient to develop the theory using the generalized coordinates ($\vec{X}$, $\xi$, $\vec{u}$) which include the reaction coordinate.
The transformation to these coordinates has non-zero Jacobian determinant $J(\vec{q}) = J(\vec{X}, \xi, \vec{u})$; since the unconstrained beads are not transformed ($\vec{u} = \vec{q}^{(2)}, \ldots, \vec{q}^{(P)}$), $J$ has no dependence on their values and we write simply $J(\vec{X}, \xi)$, noting that it has the same value as the Jacobian determinant of the transformation on just the first bead.

By taking the time derivative of the explicit constraint equation $\xi(\vec{q}^{(1)}) = \xi\st$, we find the implicit velocity constraint $\dot{\xi}(\vec{q}^{(1)}) = 0$, which prevents us from using Eq.~\eqref{eq:Z-p} directly to obtain the necessary phase space integral.
Instead, we find that the appropriate expression is
\begin{subequations}
\begin{align}
	\label{eq:Z-p-constrained}
	& \int\! \dd{\vec{p}_{\vec{U}}} \, e^{-\frac{\beta}{2} \vec{p}_{\vec{U}} \cdot \vec{A}\inv \cdot \vec{p}_{\vec{U}}}
	\notag \\
	&= \left( \frac{2 \pi}{\beta} \right)^{\frac{P f - 1}{2}} \abs{\vec{A}}^\frac{1}{2}
	= \left( \frac{2 \pi}{\beta} \right)^{\frac{P f - 1}{2}} \abs{\vec{\Gamma}}^\frac{1}{2} \, Z_\xi^\frac{1}{2} \\
	&= \left( \frac{2 \pi}{\beta} \right)^{\frac{P f - 1}{2}} \abs{\vec{M}}^\frac{1}{2} \, \abs{J(\vec{X}, \xi\st)} \, Z_\xi(\vec{X}, \xi\st)^\frac{1}{2},
\end{align}
\end{subequations}
where $\vec{p}_{\vec{U}}$ are the momenta conjugate to all the unconstrained bead coordinates $\vec{U} = \vec{X}, \vec{u}$,
\begin{align}
	\vec{\Gamma}
	&= \vec{J}\trans \cdot \vec{M} \cdot \vec{J}
\end{align}
is the generalized mass matrix, $\vec{J}$ is the Jacobian matrix (whose determinant is $J$), $\vec{A}$ is $\vec{\Gamma}$ without the row and column corresponding to $\xi$, and
\begin{align}
	\label{eq:Z-xi}
	Z_\xi
	&= \grad \xi \cdot \vec{M}\inv \cdot \grad \xi
	= \grad_1 \xi \cdot \vec{M}_1\inv \cdot \grad_1 \xi
\end{align}
is the element of $\vec{\Gamma}\inv$ at the row and column corresponding to $\xi$.
The requisite determinant identity is proved in Appendix~\ref{sec:determinant-identity}.

Therefore, we have
\begin{align}
	-\tilde{\beta} A'(\xi\st)
	&= \frac{
			\int\! \dd{\vec{p}_{\vec{U}}} \int\! \dd{\vec{U}} \, e^{-\beta H\cl\constr(\vec{p}_{\vec{U}}, \vec{U}; \xi\st)} Z_\xi^{-\frac{1}{2}} \estim_i(\vec{U}, \xi\st)
		}{
			\int\! \dd{\vec{p}_{\vec{U}}} \int\! \dd{\vec{U}} \, e^{-\beta H\cl\constr(\vec{p}_{\vec{U}}, \vec{U}; \xi\st)} Z_\xi^{-\frac{1}{2}}
		},
\end{align}
where the constrained classical Hamiltonian
\begin{align}
	H\cl\constr(\vec{p}_{\vec{U}}, \vec{U}; \xi\st)
	&= \frac{1}{2} \vec{p}_{\vec{U}} \cdot \vec{A}\inv \cdot \vec{p}_{\vec{U}} + \frac{\tilde{\beta}}{\beta} V\cl(\vec{U}, \xi\st)
\end{align}
can be obtained from Eq.~\eqref{eq:H-cl} by setting $\dot{\xi} = 0$.
We may write the above expression in terms of averages over constrained molecular dynamics simulations as
\begin{align}
	\label{eq:fixman}
	-\tilde{\beta} A'(\xi\st)
	&= \frac{
			\ev*{Z_\xi^{-\frac{1}{2}} \estim_i}_{\xi\st}
		}{
			\ev*{Z_\xi^{-\frac{1}{2}}}_{\xi\st}
		},
\end{align}
where the dependence of the constrained momentum partition function on the coordinates has given rise to the Fixman correction.\cite{fixman1978simulation,den2013revisiting}
These constrained molecular dynamics simulations are in practice carried out in Cartesian coordinates with the constraint enforced via the standard method of Lagrange multipliers.

\subsection{Constrained OBABO (c-OBABO) integrator}

The method described in Sec.~\ref{sec:constrained-hamiltonian} is applicable to constrained Hamiltonian systems, but unfortunately, the simulations of interest are to be run at constant temperature using a Langevin thermostat.
However, as Leli\`{e}vre \textit{et al.} have shown, the thermostat step may be adjusted in exactly the same manner for Langevin dynamics.\cite{lelievre2010free,lelievre2012langevin}
This allows us to transform the PILE (OBABO) integrator into the constrained version, c-OBABO.
Although for our present purposes (computing the quantum PMF), we only require a single constraint on one bead, we derive here a more general integrator that supports multiple independent holonomic constraints on all beads.

Consider $P$ functions $\vec{\xi}_j(\vec{q}^{(j)})$ and constant vectors $\vec{z}_j$ (not necessarily of the same length for each bead), making up the constraint equations $\vec{\xi}_j(\vec{q}^{(j)}) = \vec{z}_j$.
The implicit velocity constraints, when expressed in momentum form, are $\grad_j \vec{\xi}_j(\vec{q}^{(j)})\trans \cdot \vec{M}_1\inv \cdot \vec{p}^{(j)} = \vec{0}$.
Upon adding these constraints to the OBABO integrator and removing redundant constraint steps, we obtain the c-OBABO integrator for PIMD:
\begin{subequations}\small
\label{eq:cobabo}
\begin{align}
	\rotatebox[origin=c]{90}{O$_1$} &\begin{dcases}
		\vec{P}^{(k)}
		\leftarrow \textstyle\sum_{j=1}^P \vec{p}^{(j)} C_{j k} \\[1mm]
		\bar{\vec{P}}^{(k)}
		\leftarrow T_k(\dt/2) \vec{P}^{(k)} + U_k(\dt/2) \vec{M}_1^{\frac{1}{2}} \vec{\eta} \\[1 mm]
		\bar{\vec{p}}^{(j)}
		\leftarrow \textstyle\sum_{k=0}^{P-1} C_{j k} \bar{\vec{P}}^{(k)}
	\end{dcases} \\
	\rotatebox[origin=c]{90}{B$_2$} &\begin{dcases}
		\bar{\vec{p}}^{(j)}
		\leftarrow \vec{p}^{(j)} + \frac{\dt \tilde{\beta}}{2 P \beta} \vec{F}(\vec{q}^{(j)})
	\end{dcases} \\
	\rotatebox[origin=c]{90}{C$_2$} &\begin{dcases}
		\bar{\vec{p}}^{(j)}
		\leftarrow \vec{p}^{(j)} + \grad_j \vec{\xi}_j(\vec{q}^{(j)}) \cdot \vec{\Lambda}_j \\[1mm]
		\grad_j \vec{\xi}_j(\vec{q}^{(j)})\trans \cdot \vec{M}_1\inv \cdot \bar{\vec{p}}^{(j)} = \vec{0}
	\end{dcases}
\end{align}
\begin{align}
	\label{eq:subint-ho-c}
	\rotatebox[origin=c]{90}{\~{A}$_3$} &\begin{dcases}
		\vec{P}^{(k)}
		\leftarrow \textstyle\sum_{j=1}^P \vec{p}^{(j)} C_{j k} \sep
		\vec{Q}^{(k)}
		\leftarrow \textstyle\sum_{j=1}^P \vec{q}^{(j)} C_{j k} \\[1mm]
		\bar{\vec{P}}^{(k)}
		\leftarrow S^{\vec{P} \vec{P}}_k(\dt)               \vec{P}^{(k)} + S^{\vec{P} \vec{Q}}_k(\dt) \vec{M}_1 \vec{Q}^{(k)} \\
		\bar{\vec{Q}}^{(k)}
		\leftarrow S^{\vec{Q} \vec{P}}_k(\dt) \vec{M}_1\inv \vec{P}^{(k)} + S^{\vec{Q} \vec{Q}}_k(\dt)           \vec{Q}^{(k)} \\[1mm]
		\bar{\vec{p}}^{(j)}
		\leftarrow \textstyle\sum_{k=0}^{P-1} C_{j k} \bar{\vec{P}}^{(k)}
		\\ \qquad\qquad
			+ \textstyle\sum_{\ell=1}^P \tilde{S}^{\vec{P} \vec{P}}_{j \ell}(\dt) \grad_\ell \vec{\xi}_\ell(\vec{q}^{(\ell)}) \cdot \vec{\Lambda}_\ell \\
		\bar{\vec{q}}^{(j)}
		\leftarrow \textstyle\sum_{k=0}^{P-1} C_{j k} \bar{\vec{Q}}^{(k)}
		\\ \qquad\qquad
			+ \textstyle\sum_{\ell=1}^P \tilde{S}^{\vec{Q} \vec{P}}_{j \ell}(\dt) \vec{M}_1\inv \grad_\ell \vec{\xi}_\ell(\vec{q}^{(\ell)}) \cdot \vec{\Lambda}_\ell \\[1mm]
		\vec{\xi}_j(\bar{\vec{q}}^{(j)}) = \vec{z}_j
	\end{dcases} \\
	\rotatebox[origin=c]{90}{B$_4$} &\begin{dcases}
		\bar{\vec{p}}^{(j)}
		\leftarrow \vec{p}^{(j)} + \frac{\dt \tilde{\beta}}{2 P \beta} \vec{F}(\vec{q}^{(j)})
	\end{dcases} \\
	\rotatebox[origin=c]{90}{C$_4$} &\begin{dcases}
		\bar{\vec{p}}^{(j)}
		\leftarrow \vec{p}^{(j)} + \grad_j \vec{\xi}_j(\vec{q}^{(j)}) \cdot \vec{\Lambda}_j \\[1mm]
		\grad_j \vec{\xi}_j(\vec{q}^{(j)})\trans \cdot \vec{M}_1\inv \cdot \bar{\vec{p}}^{(j)} = \vec{0}
	\end{dcases} \\
	\rotatebox[origin=c]{90}{O$_5$} &\begin{dcases}
		\vec{P}^{(k)}
		\leftarrow \textstyle\sum_{j=1}^P \vec{p}^{(j)} C_{j k} \\[1mm]
		\bar{\vec{P}}^{(k)}
		\leftarrow T_k(\dt/2) \vec{P}^{(k)} + U_k(\dt/2) \vec{M}_1^{\frac{1}{2}} \vec{\eta} \\[1 mm]
		\bar{\vec{p}}^{(j)}
		\leftarrow \textstyle\sum_{k=0}^{P-1} C_{j k} \bar{\vec{P}}^{(k)}
	\end{dcases}
\end{align}
\begin{align}
	\rotatebox[origin=c]{90}{C$_5$} &\begin{dcases}
		\bar{\vec{p}}^{(j)}
		\leftarrow \vec{p}^{(j)} + \grad_j \vec{\xi}_j(\vec{q}^{(j)}) \cdot \vec{\Lambda}_j \\[1mm]
		\grad_j \vec{\xi}_j(\vec{q}^{(j)})\trans \cdot \vec{M}_1\inv \cdot \bar{\vec{p}}^{(j)} = \vec{0},
	\end{dcases}
\end{align}
\end{subequations}
where
\begin{align}
	\tilde{S}_{j \ell}^{\vec{X} \vec{X}}(\dt)
	&= \sum_{k=0}^{P-1} C_{j k} S^{\vec{X} \vec{X}}_k(\dt) C_{\ell k}
\end{align}
are the normal mode propagation coefficients transformed to real space.
For the velocity constraints, the Lagrange multipliers may be computed directly as
\begin{align}
	\vec{\Lambda}_j
	&= -\left( \grad_j \vec{\xi}_j(\vec{q}^{(j)})\trans \cdot \vec{M}_1\inv \cdot \grad_j \vec{\xi}_j(\vec{q}^{(j)}) \right)\inv
	\notag \\ &\qquad\qquad
			\cdot \grad_j \vec{\xi}_j(\vec{q}^{(j)})\trans \cdot \vec{M}_1\inv \cdot \vec{p}^{(j)},
\end{align}
without an iterative scheme.

Because both the normal mode transformations and the harmonic oscillator equations of motion are linear, in going from A in Eq.~\eqref{eq:subint-ho} to \~{A} in Eq.~\eqref{eq:subint-ho-c}, the constraint force was threaded through the sub-integrator, and is applied only at the very end of \~{A}, after the inverse normal mode transformation.
Although all the beads are coupled by the constraint and the Lagrange multipliers cannot be obtained directly in the general case, in Sec.~\ref{sec:exact-constraint} we show that this form can enable direct evaluation of the Lagrange multiplier when only one constraint is needed.
Additionally, because the constraints couple the normal modes even in the absence of interactions, the standard derivation for the optimal friction of a thermostatted harmonic oscillator is not applicable.
Despite this, we find that using the unmodified friction values from the PILE ($\gamma_k = 2 \omega_k$ for $k \ge 1$) is a valid strategy in practice.

\subsection{Constrained BAOAB (c-BAOAB) integrator}

It has been demonstrated that the alternate integrator step order BAOAB may result in a smaller time step error for PIMD.\cite{liu2016simple}
While Leli\`{e}vre \textit{et al.} only consider the ``side'' scheme (which places the thermostat on the outer sides of the integrator), the feasibility and benefits of the ``middle'' scheme (which has the thermostat centered in the integrator) have also been recently established for molecular dynamics with holonomic constraints.\cite{zhang2019unified}
The c-BAOAB integrator for PIMD may therefore be written as the following arrangement of sub-integrators and constraint steps:
\begin{subequations}\small
\label{eq:cbaoab}
\begin{align}
	\rotatebox[origin=c]{90}{B$_1$} &\begin{dcases}
		\bar{\vec{p}}^{(j)}
		\leftarrow \vec{p}^{(j)} + \frac{\dt \tilde{\beta}}{2 P \beta} \vec{F}(\vec{q}^{(j)})
	\end{dcases} \\
	\rotatebox[origin=c]{90}{C$_1$} &\begin{dcases}
		\bar{\vec{p}}^{(j)}
		\leftarrow \vec{p}^{(j)} + \grad_j \vec{\xi}_j(\vec{q}^{(j)}) \cdot \vec{\Lambda}_j \\[1mm]
		\grad_j \vec{\xi}_j(\vec{q}^{(j)})\trans \cdot \vec{M}_1\inv \cdot \bar{\vec{p}}^{(j)} = \vec{0}
	\end{dcases} \\
	\rotatebox[origin=c]{90}{\~{A}$_2$} &\begin{dcases}
		\vec{P}^{(k)}
		\leftarrow \textstyle\sum_{j=1}^P \vec{p}^{(j)} C_{j k} \sep
		\vec{Q}^{(k)}
		\leftarrow \textstyle\sum_{j=1}^P \vec{q}^{(j)} C_{j k} \\[1mm]
		\bar{\vec{P}}^{(k)}
		\leftarrow S^{\vec{P} \vec{P}}_k(\dt/2)               \vec{P}^{(k)} + S^{\vec{P} \vec{Q}}_k(\dt/2) \vec{M}_1 \vec{Q}^{(k)} \\
		\bar{\vec{Q}}^{(k)}
		\leftarrow S^{\vec{Q} \vec{P}}_k(\dt/2) \vec{M}_1\inv \vec{P}^{(k)} + S^{\vec{Q} \vec{Q}}_k(\dt/2)           \vec{Q}^{(k)} \\[1mm]
		\bar{\vec{p}}^{(j)}
		\leftarrow \textstyle\sum_{k=0}^{P-1} C_{j k} \bar{\vec{P}}^{(k)}
		\\ \qquad\qquad
			+ \textstyle\sum_{\ell=1}^P \tilde{S}^{\vec{P} \vec{P}}_{j \ell}(\dt/2) \grad_\ell \vec{\xi}_\ell(\vec{q}^{(\ell)}) \cdot \vec{\Lambda}_\ell \\
		\bar{\vec{q}}^{(j)}
		\leftarrow \textstyle\sum_{k=0}^{P-1} C_{j k} \bar{\vec{Q}}^{(k)}
		\\ \qquad\qquad
			+ \textstyle\sum_{\ell=1}^P \tilde{S}^{\vec{Q} \vec{P}}_{j \ell}(\dt/2) \vec{M}_1\inv \grad_\ell \vec{\xi}_\ell(\vec{q}^{(\ell)}) \cdot \vec{\Lambda}_\ell \\[1mm]
		\vec{\xi}_j(\bar{\vec{q}}^{(j)}) = \vec{z}_j
	\end{dcases} \\
	\rotatebox[origin=c]{90}{C$_2$} &\begin{dcases}
		\bar{\vec{p}}^{(j)}
		\leftarrow \vec{p}^{(j)} + \grad_j \vec{\xi}_j(\vec{q}^{(j)}) \cdot \vec{\Lambda}_j \\[1mm]
		\grad_j \vec{\xi}_j(\vec{q}^{(j)})\trans \cdot \vec{M}_1\inv \cdot \bar{\vec{p}}^{(j)} = \vec{0}
	\end{dcases} \\
	\rotatebox[origin=c]{90}{O$_3$} &\begin{dcases}
		\vec{P}^{(k)}
		\leftarrow \textstyle\sum_{j=1}^P \vec{p}^{(j)} C_{j k} \\[1mm]
		\bar{\vec{P}}^{(k)}
		\leftarrow T_k(\dt) \vec{P}^{(k)} + U_k(\dt) \vec{M}_1^{\frac{1}{2}} \vec{\eta} \\[1 mm]
		\bar{\vec{p}}^{(j)}
		\leftarrow \textstyle\sum_{k=0}^{P-1} C_{j k} \bar{\vec{P}}^{(k)}
	\end{dcases} \\
	\rotatebox[origin=c]{90}{C$_3$} &\begin{dcases}
		\bar{\vec{p}}^{(j)}
		\leftarrow \vec{p}^{(j)} + \grad_j \vec{\xi}_j(\vec{q}^{(j)}) \cdot \vec{\Lambda}_j \\[1mm]
		\grad_j \vec{\xi}_j(\vec{q}^{(j)})\trans \cdot \vec{M}_1\inv \cdot \bar{\vec{p}}^{(j)} = \vec{0}
	\end{dcases} \\
	\rotatebox[origin=c]{90}{\~{A}$_4$} &\begin{dcases}
		\vec{P}^{(k)}
		\leftarrow \textstyle\sum_{j=1}^P \vec{p}^{(j)} C_{j k} \sep
		\vec{Q}^{(k)}
		\leftarrow \textstyle\sum_{j=1}^P \vec{q}^{(j)} C_{j k} \\[1mm]
		\bar{\vec{P}}^{(k)}
		\leftarrow S^{\vec{P} \vec{P}}_k(\dt/2)               \vec{P}^{(k)} + S^{\vec{P} \vec{Q}}_k(\dt/2) \vec{M}_1 \vec{Q}^{(k)} \\
		\bar{\vec{Q}}^{(k)}
		\leftarrow S^{\vec{Q} \vec{P}}_k(\dt/2) \vec{M}_1\inv \vec{P}^{(k)} + S^{\vec{Q} \vec{Q}}_k(\dt/2)           \vec{Q}^{(k)} \\[1mm]
		\bar{\vec{p}}^{(j)}
		\leftarrow \textstyle\sum_{k=0}^{P-1} C_{j k} \bar{\vec{P}}^{(k)}
		\\ \qquad\qquad
			+ \textstyle\sum_{\ell=1}^P \tilde{S}^{\vec{P} \vec{P}}_{j \ell}(\dt/2) \grad_\ell \vec{\xi}_\ell(\vec{q}^{(\ell)}) \cdot \vec{\Lambda}_\ell \\
		\bar{\vec{q}}^{(j)}
		\leftarrow \textstyle\sum_{k=0}^{P-1} C_{j k} \bar{\vec{Q}}^{(k)}
		\\ \qquad\qquad
			+ \textstyle\sum_{\ell=1}^P \tilde{S}^{\vec{Q} \vec{P}}_{j \ell}(\dt/2) \vec{M}_1\inv \grad_\ell \vec{\xi}_\ell(\vec{q}^{(\ell)}) \cdot \vec{\Lambda}_\ell \\[1mm]
		\vec{\xi}_j(\bar{\vec{q}}^{(j)}) = \vec{z}_j
	\end{dcases} \\
	\rotatebox[origin=c]{90}{B$_5$} &\begin{dcases}
		\bar{\vec{p}}^{(j)}
		\leftarrow \vec{p}^{(j)} + \frac{\dt \tilde{\beta}}{2 P \beta} \vec{F}(\vec{q}^{(j)})
	\end{dcases}
\end{align}
\begin{align}
	\rotatebox[origin=c]{90}{C$_5$} &\begin{dcases}
		\bar{\vec{p}}^{(j)}
		\leftarrow \vec{p}^{(j)} + \grad_j \vec{\xi}_j(\vec{q}^{(j)}) \cdot \vec{\Lambda}_j \\[1mm]
		\grad_j \vec{\xi}_j(\vec{q}^{(j)})\trans \cdot \vec{M}_1\inv \cdot \bar{\vec{p}}^{(j)} = \vec{0}.
	\end{dcases}
\end{align}
\end{subequations}

Our rather unsophisticated implementation\cite{watermeanmeanforcejl} of c-BAOAB suffers an increase in run time on the order of \SI{10}{\percent} compared to c-OBABO due to the additional work required.
Despite this, as we show in Sec.~\ref{sec:results}, the former may still outperform the latter, since it can allow a much larger time step to be used.

As pointed out in Ref.~\onlinecite{zhang2019unified}, the momentum distribution sampled by integrators which do not have the thermostat as the final step will typically deviate from the expected Maxwell--Boltzmann distribution.
In Appendix~\ref{sec:temperature}, we briefly explore the consequences of this with respect to the observed temperature.

\section{Exact relations for a center of mass distance constraint}
\label{sec:exact-constraint}

When the reaction coordinate $\xi$ is the radial separation between two centers of mass, several useful expressions may be derived.
From this point onward, we work explicitly with $N$ particles in 3-dimensional space, with coordinates $\vec{x}_i$ (where $\vec{x}$ describes all the degrees of freedom at the first bead, and can be thought of as a more structured reinterpretation of $\vec{q}^{(1)}$).
The convex sums
\begin{subequations}
\begin{align}
	\vec{x}_\alpha
	&= \sum_{i=1}^{N_\alpha} \frac{m_{\alpha i}}{m_\alpha} \vec{x}_{\alpha i} \\
\intertext{and}
	\vec{x}_\beta
	&= \sum_{i=1}^{N_\beta} \frac{m_{\beta i}}{m_\beta} \vec{x}_{\beta i}
\end{align}
\end{subequations}
are the centers of mass of $\alpha$ and $\beta$ (which we require to be non-empty and disjoint), where $m_{\alpha i}$ is the mass corresponding to $\vec{x}_{\alpha i}$ and
\begin{align}
	m_\alpha
	&= \sum_{i=1}^{N_\alpha} m_{\alpha i},
\end{align}
and likewise for $\beta$.
We also use $\vec{x}_{\gamma i}$ to refer to the remaining particles, if any, which do not participate in either center of mass, with $N_\alpha + N_\beta + N_\gamma = N$.

We constrain the distance between the centers of mass as
\begin{align}
	\xi(\vec{x})
	&= \abs{\vec{r}}
	= \abs{\vec{x}_\alpha - \vec{x}_\beta}
	= z,
\end{align}
for an arbitrary positive $z$.
The derivatives of this function are
\begin{subequations}
\begin{align}
	\grad_{\alpha i} \xi(\vec{x})
	&= \phantom{-}\frac{m_{\alpha i} \vec{r}}{m_\alpha \xi(\vec{x})}, \\
	\grad_{\beta i} \xi(\vec{x})
	&= -\frac{m_{\beta i} \vec{r}}{m_\beta \xi(\vec{x})},
\intertext{and}
	\grad_{\gamma i} \xi(\vec{x})
	&= 0.
\end{align}
\end{subequations}
Immediately, it follows that Eq.~\eqref{eq:Z-xi} evaluates to
\begin{subequations}
\begin{align}
	Z_\xi(\vec{x})
	&= \sum_{i=1}^{N_\alpha} \frac{1}{m_{\alpha i}} \abs{\grad_{\alpha i} \xi(\vec{x})}^2
		+ \sum_{i=1}^{N_\beta} \frac{1}{m_{\beta i}} \abs{\grad_{\beta i} \xi(\vec{x})}^2 \\
	&= \frac{1}{m_\alpha} + \frac{1}{m_\beta}
	= \frac{1}{\mu_{\alpha \beta}},
\end{align}
\end{subequations}
which is a constant (with $\mu_{\alpha \beta}$ being the reduced mass), so the Fixman correction cancels from Eq.~\eqref{eq:fixman} leaving us with just
\begin{align}
	-\tilde{\beta} A'(\xi\st)
	&= \ev*{\estim_i}_{\xi\st}.
\end{align}

\subsection{Lagrange multiplier}

It is also straightforward in this circumstance to obtain a closed-form solution for the Lagrange multiplier $\lambda$ for the constrained harmonic oscillator propagation step \~{A}.
Because the momentum perturbation was propagated through the linear equations of motion, we know that the equation to be satisfied is
\begin{align}
	\xi\!\left( \bar{\vec{x}} + \lambda \tilde{S}^{\vec{Q} \vec{P}}_{1 1} \vec{M}_1\inv \grad_1 \xi(\vec{x}) \right) = z,
\end{align}
where $\vec{x}$ is the value of $\vec{q}^{(1)}$ before the propagation, and $\bar{\vec{x}}$ is its value after propagation by the unconstrained step A.
After some algebraic manipulations, we get
\begin{align}
	\abs{\left( 1 + \frac{\lambda \tilde{S}^{\vec{Q} \vec{P}}_{1 1}}{\mu_{\alpha \beta} \xi(\vec{x})} \right) \vec{r} + \Delta \vec{r}}
	&= z,
\end{align}
where $\Delta \vec{r} = \bar{\vec{r}} - \vec{r}$ is the change in relative position of the centers of mass due to the unconstrained step.
We assume that $\tilde{S}^{\vec{Q} \vec{P}}_{1 1}$ is not zero.

From geometric considerations, we see that solutions exist only when the component of $\Delta \vec{r}$ orthogonal to $\vec{r}$ does not surpass $z$:
\begin{align}
	\label{eq:constraint-bound}
	\abs{\Delta \vec{r}}^2 - (\Delta \vec{r} \cdot \vec{\hat{r}})^2
	&\le z^2,
\end{align}
where $\vec{\hat{r}} = \vec{r} / \abs{\vec{r}}$ is the unit vector in the direction of $\vec{r}$.
When this condition is met, the Lagrange multiplier is given by
\begin{align}
	\label{eq:com-lagrange}
	\lambda
	&= -\frac{\mu_{\alpha \beta}}{\tilde{S}^{\vec{Q} \vec{P}}_{1 1}} \left[
			\abs{\vec{r}}
			+ \Delta \vec{r} \cdot \vec{\hat{r}}
			- \sqrt{z^2 - \left( \abs{\Delta \vec{r}}^2 - (\Delta \vec{r} \cdot \vec{\hat{r}})^2 \right)}
		\right],
\end{align}
which may be computed directly for very little cost.
When the square root is not zero, there is another solution for $\lambda$, which has the root being added rather than subtracted, but this corresponds to the interchange of the two centers of mass; we assume that the time step is sufficiently small that this will never be the desired outcome.

\begin{figure}
	\includegraphics{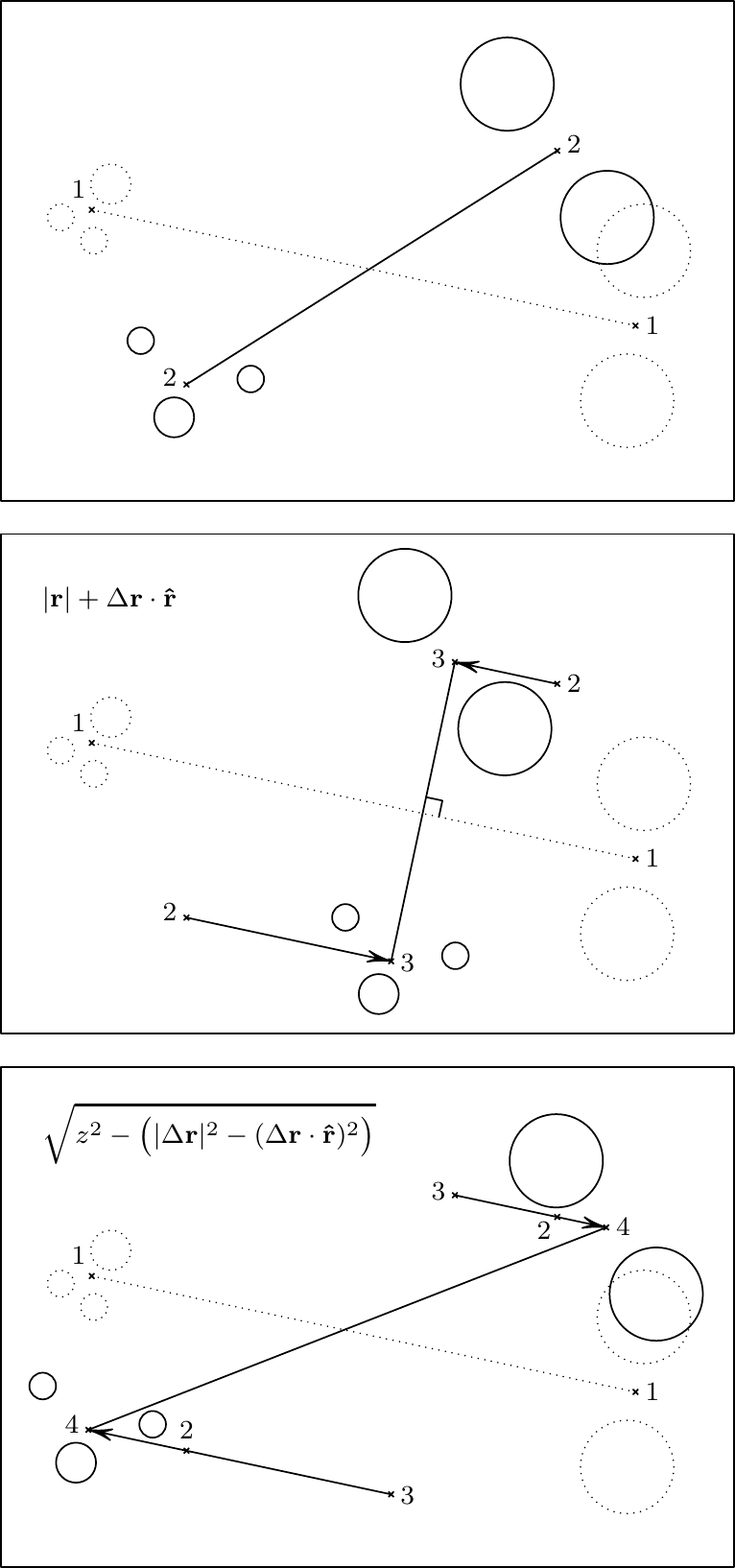}
	\caption{
		Schematic depiction of the shifts in Eq.~\eqref{eq:constraint-shifts} for a system with five particles split into two centers of mass.
		Each panel shows the original configuration using dotted circles; the line segment connecting their centers of mass (1) has length $z$ and is parallel to $\vec{\hat{r}}$.
		In the top panel, the particles are at their new locations after propagation by the unconstrained step A; their centers of mass (2) are closer than $z$, violating the constraint.
		In the middle panel, the positions have been adjusted by the first two terms of the shifts; the line segment connecting their centers of mass (3) is orthogonal to $\vec{\hat{r}}$.
		In the bottom panel, the particles are at their final locations; the distance between their centers of mass (4) is $z$, satisfying the constraint.
	}
	\label{fig:multiplier}
\end{figure}

The interpretation of the expression for $\lambda$ is simple, as illustrated in Fig.~\ref{fig:multiplier}.
The resulting shifts of the particles in $\alpha$ and $\beta$ are, respectively,
\begin{subequations}
\label{eq:constraint-shifts}
\begin{align}
	-\frac{m_\beta \vec{\hat{r}}}{m_\alpha + m_\beta} \left[
			\abs{\vec{r}}
			+ \Delta \vec{r} \cdot \vec{\hat{r}}
			- \sqrt{z^2 - \left( \abs{\Delta \vec{r}}^2 - (\Delta \vec{r} \cdot \vec{\hat{r}})^2 \right)}
		\right]
\end{align}
and
\begin{align}
	\frac{m_\alpha \vec{\hat{r}}}{m_\alpha + m_\beta} \left[
			\abs{\vec{r}}
			+ \Delta \vec{r} \cdot \vec{\hat{r}}
			- \sqrt{z^2 - \left( \abs{\Delta \vec{r}}^2 - (\Delta \vec{r} \cdot \vec{\hat{r}})^2 \right)}
		\right].
\end{align}
\end{subequations}
The first two terms of each are responsible for completely removing any separation between the centers of mass along $\vec{r}$: the first term takes care of the original vector, while the second term handles the component of $\Delta \vec{r}$ that is parallel to $\vec{r}$.
The prefactors ensure that the groups move toward one another, and that the lighter group moves farther.
The square root term then restores some of the separation along $\vec{r}$, with the exact amount being determined by the excess allowance from Eq.~\eqref{eq:constraint-bound}.

It is important to note that Eq.~\eqref{eq:com-lagrange} is not an equation of motion for the constraint.
Rather, this is an exact solution to the Lagrange multiplier optimization, equivalent to one that would be obtained by an iterative scheme.
As such, it is not susceptible to numerical drift of the constrained coordinate.

\subsection{Derivatives for estimators}
\label{sec:estimator-derivatives}

Evaluation of the estimators $\estim_1$ and $\estim_2$ requires knowledge of $\dlogJ{\vec{x}}$ and $\pdv{\vec{x}}{\xi}$, which we find by performing several coordinate transformations.
We begin by transforming $\vec{x}$ into the Jacobi coordinates
\begin{align}
	\vec{y}_{\alpha \ell}
	&= \sum_{i=1}^\ell \frac{m_{\alpha i}}{m_\alpha^{1 \to \ell}} \vec{x}_{\alpha i}
			- \begin{dcases*}
					\vec{x}_{\alpha, \ell+1} & if $1 \le \ell \le N_\alpha - 1$ \\
					0 & if $\ell = N_\alpha$,
				\end{dcases*}
\end{align}
with
\begin{align}
	m_\alpha^{1 \to \ell}
	&= \sum_{i=1}^\ell m_{\alpha i},
\end{align}
and similarly for $\beta$.
The remaining coordinates $\vec{x}_{\gamma i}$ are left unmodified.
As shown in Appendix~\ref{sec:jacobi-jacobian}, this transformation has unit Jacobian determinant.

The coordinates $\vec{y}_{\alpha N_\alpha}$ and $\vec{y}_{\beta N_\beta}$ are the centers of mass $\vec{x}_\alpha$ and $\vec{x}_\beta$, so we further transform them to
\begin{subequations}
\begin{align}
	\vec{R}
	&= \frac{m_\alpha \vec{y}_{\alpha N_\alpha} + m_\beta \vec{y}_{\beta N_\beta}}{m_\alpha + m_\beta} \\
\intertext{and}
	\vec{r}
	&= \vec{y}_{\alpha N_\alpha} - \vec{y}_{\beta N_\beta},
\end{align}
\end{subequations}
and this change of variables also has unit Jacobian determinant.

Finally, the transformation of $\vec{r}$ to the spherical coordinates ($\xi$, $\cos{\theta}$, $\phi$) is known to have a Jacobian determinant whose absolute value is $\xi^2$.
The overall transformation from $\vec{x}$ to ($\vec{y}_{\alpha i}$, $\vec{y}_{\beta i}$, $\vec{R}$, $\xi$, $\cos{\theta}$, $\phi$, $\vec{x}_{\gamma i}$) therefore has $\abs{J(\vec{x})} = \xi^2$, so
\begin{align}
	\dlogJ{\vec{x}}
	&= \frac{2}{\xi}.
\end{align}
The original Cartesian coordinates may be written as
\begin{subequations}
\begin{align}
	\vec{x}_{\alpha i}
	&= \vec{R} + \frac{m_\beta \vec{r}}{m_\alpha + m_\beta}
			+ \sum_{\ell = i}^{N_\alpha - 1} \frac{m_{\alpha, \ell+1}}{m_\alpha^{1 \to \ell+1}} \vec{y}_{\alpha \ell}
	\notag \\ &\qquad\quad
			- \begin{dcases*}
					0 & if $i = 1$ \\
					\frac{m_\alpha^{1 \to i-1}}{m_\alpha^{1 \to i}} \vec{y}_{\alpha, i-1} & if $2 \le i \le N_\alpha$
				\end{dcases*}
\intertext{and}
	\vec{x}_{\beta i}
	&= \vec{R} - \frac{m_\alpha \vec{r}}{m_\alpha + m_\beta}
			+ \sum_{\ell = i}^{N_\beta - 1} \frac{m_{\beta, \ell+1}}{m_\beta^{1 \to \ell+1}} \vec{y}_{\beta \ell}
	\notag \\ &\qquad\quad
			- \begin{dcases*}
					0 & if $i = 1$ \\
					\frac{m_\beta^{1 \to i-1}}{m_\beta^{1 \to i}} \vec{y}_{\beta, i-1} & if $2 \le i \le N_\beta$.
				\end{dcases*}
\end{align}
\end{subequations}
Hence, the derivatives of $\vec{x}$ with respect to $\xi$ are
\begin{subequations}
\begin{align}
	\pdv{\vec{x}_{\alpha i}}{\xi}
	&= \phantom{-}\frac{m_\beta}{m_\alpha + m_\beta} \vec{\hat{r}}, \\
	\pdv{\vec{x}_{\beta i}}{\xi}
	&= -\frac{m_\alpha}{m_\alpha + m_\beta} \vec{\hat{r}}, \\
\intertext{and}
	\pdv{\vec{x}_{\gamma i}}{\xi}
	&= 0.
\end{align}
\end{subequations}
As observed in the shifts to enforce the constraints in Eq.~\eqref{eq:constraint-shifts}, the prefactor for the lighter group is larger in magnitude than for heavier group: the former must move faster under a changing separation distance than the latter.

\section{Results}
\label{sec:results}

To demonstrate the effectiveness of the constrained PIMD integrators, we apply them to a q-SPC/Fw\cite{paesani2006accurate} water dimer and obtain its quantum PMF as a function of the distance $\xi(\vec{x}) = \abs{\vec{x}_\alpha - \vec{x}_\beta}$ between the molecules' centers of mass $\vec{x}_\alpha$ and $\vec{x}_\beta$.
To generate reference results, we use the path integral umbrella sampling method (US/WHAM), which requires a histogram unbiasing step in order to stitch together the obtained histograms.\cite{bishop2018quantum}

\begin{figure}
	\includegraphics{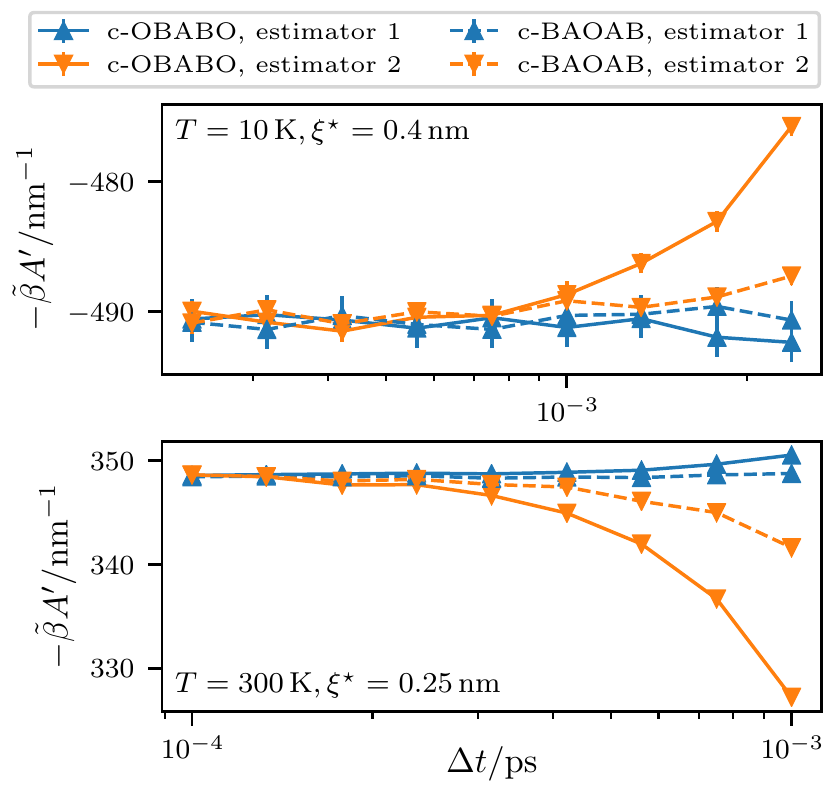}
	\caption{
		Comparison of the time step convergence of the PMF derivative of a water dimer using the integrators c-OBABO and c-BAOAB in Eqs.~\eqref{eq:cobabo} and \eqref{eq:cbaoab} at $T = \SI{10}{\kelvin}$, $\xi\st = \SI{0.4}{\nano\meter}$ (top panel) and $T = \SI{300}{\kelvin}$, $\xi\st = \SI{0.25}{\nano\meter}$ (bottom panel).
		Line segments added to guide the eye.
	}
	\label{fig:timestep}
\end{figure}

Before attempting to generate a PMF, we first identify which integrator has better time step error characteristics.
In Fig.~\ref{fig:timestep}, we show two combinations of temperature and constraint distance that have an appreciable difference in behavior between c-OBABO (solid) and c-BAOAB (dashed).
In both cases, the time step error decreases faster for the latter, which is to be expected from past work with this ordering of integrator steps.\cite{liu2016simple,zhang2019unified}
For the remainder of the calculations, we only use c-BAOAB, as it allows us to save some computational effort by using a larger time step to achieve the same level of error.

\begin{figure}
	\includegraphics{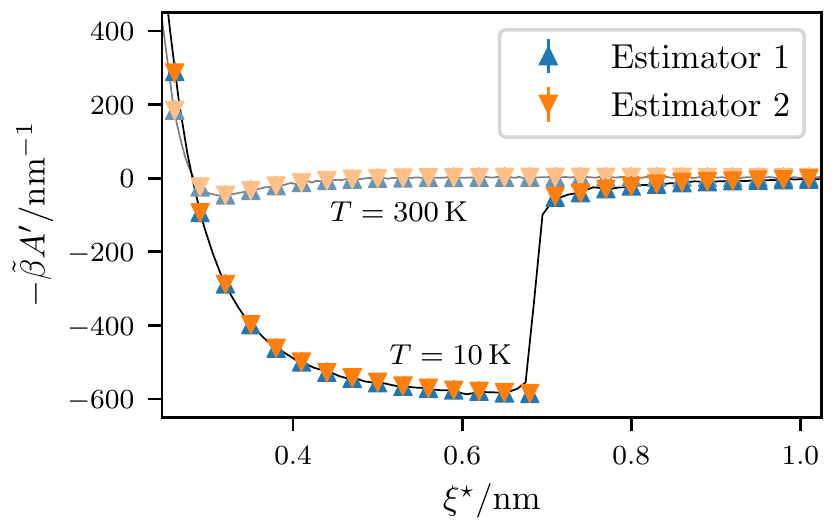}
	\caption{
		Derivative of the PMF of a water dimer at $T = \SI{300}{\kelvin}$ (top curve, less saturated) and \SI{10}{\kelvin} (bottom curve, more saturated).
		Error bars are not visible, because they are smaller than the symbols.
		The solid curves show the US/WHAM results.
	}
	\label{fig:forces}
\end{figure}

As shown in Fig.~\ref{fig:forces}, direct estimation of the derivative of the PMF is successful using the c-BAOAB integrator and both estimators.
The reference curves are obtained by numerically differentiating the US/WHAM PMF, which causes a slight wobble that is noticeable at $\SI{300}{\kelvin}$.
At $\SI{10}{\kelvin}$, the PIMD simulations show that the derivative changes rather sharply near $\SI{0.7}{\nano\meter}$, but the reference curve has rounded corners.
This region of the reaction coordinate is difficult to simulate, as the system straddles the boundary between being two largely free monomers and a strongly bound dimer.
This makes it challenging to obtain accurate histograms, and even more so for their derivatives, but we believe that the estimators for the direct calculation of the PMF derivative are not as sensitive to this sampling difficulty.
This allows us to readily estimate the derivative of the PMF at any point along the reaction coordinate from a single PIMD simulation.

\begin{figure}
	\includegraphics{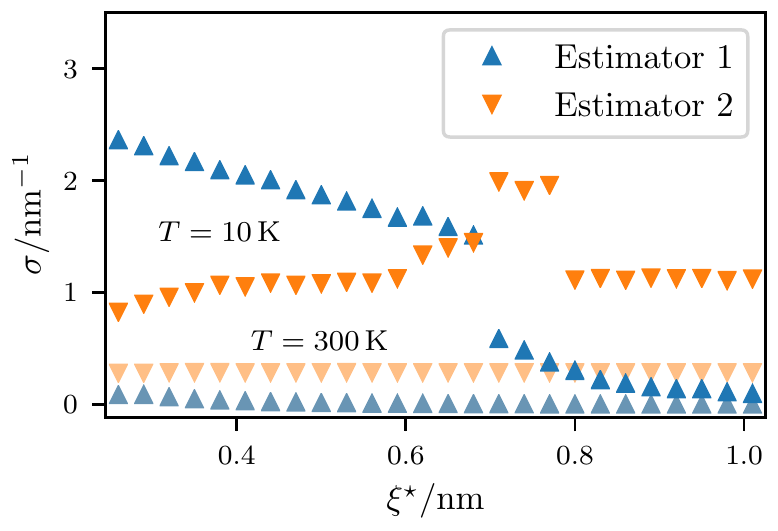}
	\caption{
		Comparison of the standard error of the mean of the PMF derivatives in Fig.~\ref{fig:forces} at $T = \SI{300}{\kelvin}$ (lower points, less saturated) and \SI{10}{\kelvin} (upper points, more saturated).
	}
	\label{fig:errors}
\end{figure}

The error bars in Fig.~\ref{fig:forces} are small enough that they cannot be seen at that scale, so we present them separately in Fig.~\ref{fig:errors}.
At $\SI{300}{\kelvin}$, the errors are relatively constant, without any interesting features.
However, the picture is not quite as simple at $\SI{10}{\kelvin}$, which has a crossing precisely in the problematic region near $\xi\st = \SI{0.7}{\nano\meter}$.
This suggests that while both sides of the identity in Eq.~(42) of Paper I converge to the same value, one may be preferable to the other depending on the circumstances.

\begin{figure}
	\includegraphics{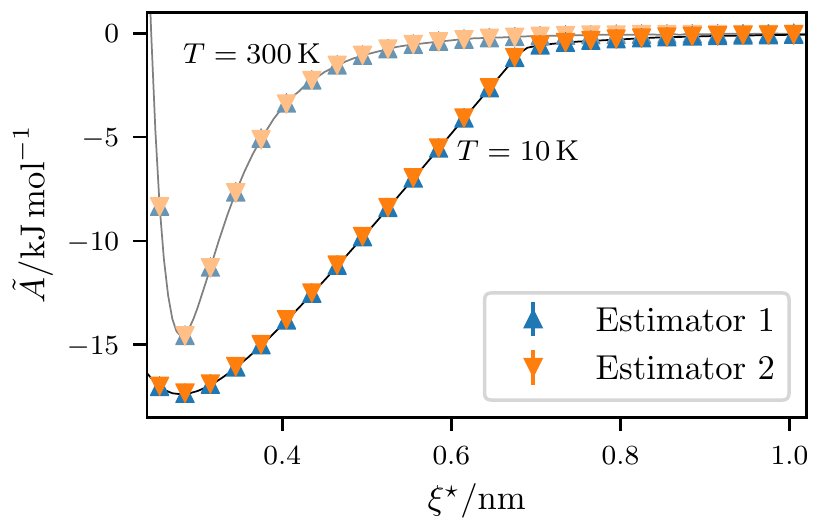}
	\caption{
		Potential of mean force of a water dimer at $T = \SI{300}{\kelvin}$ (top curve, less saturated) and \SI{10}{\kelvin} (bottom curve, more saturated).
		Error bars are not visible, because they are smaller than the symbols.
		Additional points extending to $\xi\st_0 = \SI{1.5}{\nano\meter}$ are not displayed, and the integration grid is 3 times as dense as depicted here and in Fig.~\ref{fig:forces}.
		The solid curves show the US/WHAM results.
	}
	\label{fig:pmf}
\end{figure}

Finally, we numerically integrate the obtained $-\tilde{\beta} A'$ values and plot the renormalized PMF $\tilde{A}$ in Fig.~\ref{fig:pmf}, using the same procedure as in Paper I.
There is excellent agreement with the reference results at both temperatures, suggesting that this approach of integrating the derivative of the PMF is a viable alternative to umbrella sampling and histogram unbiasing.

\section{Conclusions}
\label{sec:conclusions}

We have augmented the PILE integrator with bead-local holonomic constraints to produce the c-OBABO and c-BAOAB integrators for PIMD.
These integrators allow our estimators from Paper I to be used with molecular dynamics in addition to Monte Carlo, which greatly expands their scope of applicability.
The constrained harmonic oscillator propagation step \~{A} of these integrators has the constraint force applied at the end rather than the beginning of the step, which allows us to find an exact expression for the Lagrange multiplier for some reaction coordinates, such as the distance between two centers of mass.

Using one of the constrained integrators, we have computed the derivative of the PMF of a water dimer at \SI{10}{\kelvin} and \SI{300}{\kelvin}.
We observe that the constrained PIMD method captures the sharp step in the derivative better than umbrella sampling.
We have also been able to successfully integrate the derivative to obtain the PMF itself.

Together, these novel estimators and integrators may be utilized to find free energy differences and potentials of mean force of molecular clusters using implementations based on existing PIMD simulation software.
The conventional umbrella sampling and histogram unbiasing approach has several practical drawbacks: the user must choose a force constant for the restraint; the ends of the histogram have a tendency to be noisy, so histograms must be generated well past the region of interest; in addition to requiring a grid of points for the unbiasing step, the user must also select the locations of the restraint windows and make sure that the resulting histograms have sufficient overlap; the iterative unbiasing procedure requires a stopping criterion.
Our method, on the other hand, allows the result of each PIMD simulation to be converged independently of all the others, and there are no additional tunable parameters introduced into the simulations besides the choice of integration grid.

\begin{acknowledgments}
This research was supported by the Natural Sciences and Engineering Research Council of Canada (NSERC) (RGPIN-2016-04403), the Ontario Ministry of Research and Innovation (MRI), the Canada Research Chair program (950-231024), and the Canada Foundation for Innovation (CFI) (project No. 35232).
\end{acknowledgments}

\appendix

\section{Block matrix determinant identity}
\label{sec:determinant-identity}

Consider an invertible block matrix and its inverse with identical block structure,
\begin{align}
	\begin{pmatrix}
		\vec{A} & \vec{B} \\
		\vec{C} & \vec{D} \\
	\end{pmatrix}
	&= \begin{pmatrix}
		\vec{W} & \vec{X} \\
		\vec{Y} & \vec{Z} \\
	\end{pmatrix}\inv,
\end{align}
whose blocks $\vec{A}$ and $\vec{Z}$ are themselves invertible.
Upon combining two manifestations of the Schur complement,\cite{cottle1974manifestations}
\begin{align}
	\abs{\begin{matrix}
		\vec{A} & \vec{B} \\
		\vec{C} & \vec{D} \\
	\end{matrix}}
	&= \abs{\vec{A}} \, \abs{\vec{D} - \vec{C} \vec{A}\inv \vec{B}}
\end{align}
and
\begin{align}
	\vec{Z}\inv
	&= \vec{D} - \vec{C} \vec{A}\inv \vec{B},
\end{align}
we conclude that
\begin{align}
	\abs{\begin{matrix}
		\vec{A} & \vec{B} \\
		\vec{C} & \vec{D} \\
	\end{matrix}}
	&= \abs{\vec{A}} \, \abs{\vec{Z}}^{-1}.
\end{align}

\section{Temperature deviations}
\label{sec:temperature}

In a molecular dynamics simulation, the temperature $T$ is a parameter that controls the integrator thermostat, which tries to ensure that the momenta are distributed according to the Maxwell--Boltzmann distribution.
However, when the integrator does not end on a thermostat step, the efforts of the thermostat may be disrupted.\cite{zhang2019unified}
Since we expect that
\begin{subequations}
\begin{align}
	T
	&= \frac{\ev*{\vec{p} \cdot \vec{M}\inv \cdot \vec{p}}}{\kB P f} \\
\intertext{and}
	T
	&= \frac{\ev*{\vec{p} \cdot \vec{M}\inv \cdot \vec{p}}_{\xi\st}}{\kB (P f - 1)}
\end{align}
\end{subequations}
for unconstrained and constrained PIMD simulations, respectively, we may observe any deviations from the target distribution by estimating the simulation temperature.

\begin{figure}
	\includegraphics{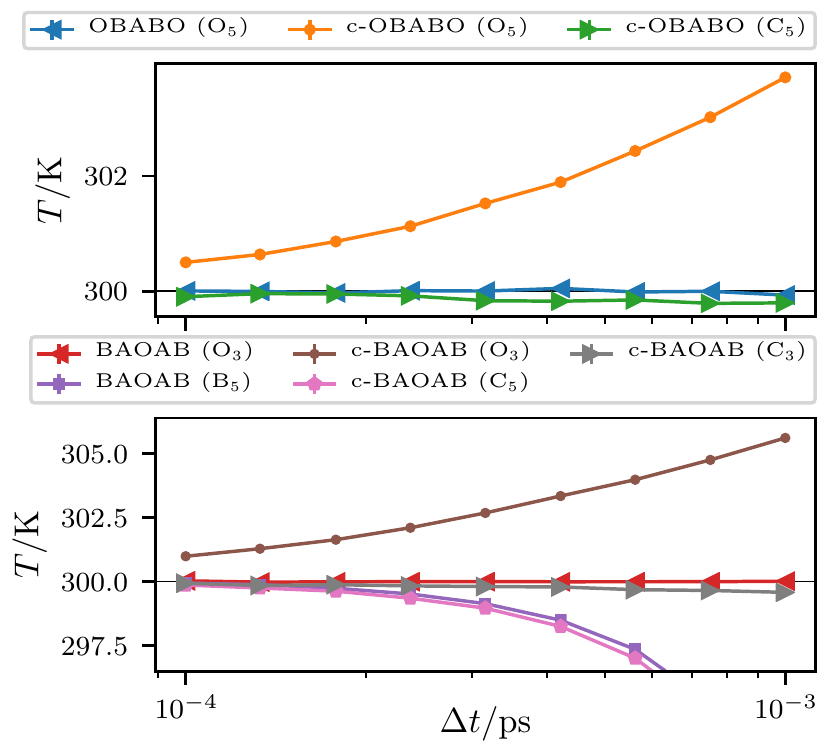}
	\caption{
		Convergence of estimated temperature with time step for several integrators and measurement points.
		Error bars are not visible, because they are smaller than the symbols.
		The momenta used to estimate the kinetic energy are taken after the integrator step indicated in parentheses.
		The horizontal line at $T = \SI{300}{\kelvin}$ marks the desired simulation temperature.
	}
	\label{fig:temperatures}
\end{figure}

The OBABO integrator (top panel of Fig.~\ref{fig:temperatures}, left-facing triangles) happens to end on a thermostat step (O$_5$), so there are no problems with the resulting temperature.
However, for the BAOAB integrator, using the end-of-integrator momentum distribution (bottom panel, squares) results in a systematic error for the temperature, as the A$_4$ and B$_5$ steps alter the momenta generated by the thermostat.
To remedy this, one can measure the temperature immediately after the thermostat step (bottom panel, left-facing triangles).
A similar issue is present for the c-BAOAB integrator (bottom panel, pentagons).

For the constrained integrators, there is an additional source of error that arises if the temperature is measured before the velocity constraint is applied.
Because the Langevin thermostat applies the random kick to all $P f$ degrees of freedom, it introduces too much energy into the system (both panels, circles).
Allowing the constraint step to project out the momentum in the constrained direction significantly reduces the magnitude of the error (both panels, right-facing triangles).
The remaining discrepancy is due to the stochastic nature of the thermostat: the instantaneous kinetic energy is not equally partitioned between the degrees of freedom, so the constraint step may remove insufficient or excess energy from the system.
As suggested in Ref.~\onlinecite{zhang2019unified}, a self-consistent iterative procedure could rectify this fully, but that is not necessary in practice.
In all cases, we observe that the error in the temperature decreases as the time step $\dt$ is made smaller, since the sources of error can contribute less of it over a more limited duration.

\section{Jacobian determinant of transformation to Jacobi coordinates}
\label{sec:jacobi-jacobian}

The transformation to Jacobi coordinates described in Sec.~\ref{sec:estimator-derivatives} couples neither $\alpha$ and $\beta$ nor the spatial degrees of freedom, and also does not involve $\gamma$, so its Jacobian matrix will be block diagonal with 7 independent blocks.
The entire $\gamma$ block is irrelevant, as its determinant is 1.
The remaining 6 blocks are identical in form, so we may scrutinize only one of them, using $x_i$ to label an arbitrary Cartesian component of either $\vec{x}_{\alpha i}$ or $\vec{x}_{\beta i}$, with corresponding masses $m_i$ and transformed coordinates $y_\ell$.

Trivially, the derivative of $y_\ell$ with respect to $x_i$ is
\begin{align}
	\pdv{y_\ell}{x_i}
	&= \begin{dcases*}
			\frac{m_i}{m^{1 \to \ell}} & if $i \le \ell$ \\
			-1 & if $i = \ell + 1$ \\
			0 & otherwise,
		\end{dcases*}
\end{align}
so the Jacobian matrix $\vec{J}$ with elements
\begin{align}
	J_{\ell i}
	&= \pdv{y_\ell}{x_i}
\end{align}
is a lower Hessenberg matrix (all elements above the superdiagonal $i = \ell + 1$ are zero).
Thus, we may use a recurrence relation to find $\abs{\vec{J}}$.\cite{cahill2002fibonacci}

In the following, $\vec{J}^{(n)}$ denotes the $n \times n$ leading principal submatrix of $\vec{J}$, and we see that $\abs*{\vec{J}^{(0)}} = 1$ and $\abs*{\vec{J}^{(1)}} = 1$.
For subsequent $n$,
\begin{subequations}
\begin{align}
	\abs*{\vec{J}^{(n)}}
	&= J_{n n} \abs*{\vec{J}^{(n-1)}}
		+ \sum_{i=1}^{n-1} (-1)^{n-i} J_{n i} \prod_{j=i}^{n-1} J_{j, j+1} \abs*{\vec{J}^{(j-1)}} \\
	&= \frac{m_n}{m^{1 \to n}} \abs*{\vec{J}^{(n-1)}}
		+ \sum_{i=1}^{n-1} \frac{m_i}{m^{1 \to n}} \prod_{j=i}^{n-1} \abs*{\vec{J}^{(j-1)}}.
\end{align}
\end{subequations}
By induction, it stands to reason that $\abs*{\vec{J}^{(n)}} = 1$ for all $n$, and therefore $\abs*{\vec{J}} = 1$, so the entire transformation to Jacobi coordinates for both centers of mass has unit Jacobian determinant.
This is consistent with the view of Jacobi coordinates as iterated pairwise transformations to center of mass and relative distance coordinates, which individually have unit Jacobian determinant.

\end{document}